\documentclass[aps,prl,reprint,showpacs,twocolumn,longbibliography,superscriptaddress]{revtex4-1}

\usepackage{graphicx}
\usepackage{dcolumn}
\usepackage{bm}
\usepackage{braket}
\usepackage{leftidx}
\usepackage{cancel}
\usepackage{amsmath}
\usepackage{epstopdf}

\begin{document}


\title{Many-particle entanglement criterion for superradiant-like states }

\author{Mehmet Emre Tasgin}

\affiliation{Institute of Nuclear Sciences, Hacettepe University, 06800, Ankara, Turkey}
\affiliation{metasgin@hacettepe.edu.tr}





\date{\today}

\begin{abstract}
We derive a many-particle inseparability criterion for mixed states using the relation between single-mode and many-particle nonclassicalities. It works very well not only in the vicinity of the Dicke states, but also for the superposition of them: superradiant ground state of finite/infinite number of particles and time evolution of single-photon superradiance. We also obtain a criterion for ensemble-field entanglement which works fine for such kind of states. Even though the collective excitations of the many-particle system is sub-Poissonian --which results in entanglement-- the wave function displays bunching.
\end{abstract}


\maketitle

Three kinds of nonclassicalities, (i) many-particle inseparability, (ii) two-mode entanglement and (iii) single-mode nonclassicality, are intimately connected to each other~\cite{hillery2006PRA,tasgin2015HP,tasgin2015measure}. Single-mode nonclassicality can be transformed to two-mode entanglement using a beam-splitter (BS) and vice versa~\cite{Kim:02,Wang:02,Asboth:05}. Although nonclassicality can be transformed into two-mode entanglement partially~\cite{ge2015conservation,Miranowicz2016nonclassicality,Miranowicz2016interplay}, this relation can be utilized for converting two-mode entanglement witnesses~\cite{Asboth:05,Tahira:09} into single-mode nonclassicality criteria~\cite{hillery2006PRA}. Such a relation is also encountered between two-mode entanglement and many-particle inseparability in Ref.~\cite{dalton2014NJP}. It is shown that spin-squeezing criterion~\cite{sorensen2001Nature} (many-particle inseparability) cannot be satisfied unless the two-modes describing this N-particle two-level system is entangled. 

A link between the many-particle inseparability and the single-mode nonclassicality shows up after one realizes the following connection. Atomic coherent states (ACSs) ---separable symmetric many-particle states~\cite{arecchi1972ACS,MandelWolfbook}--- converges to coherent (classical) states of light in the $N\to\infty$ limit~\cite{radcliffe1971JPhysA,klauder1985applications}. Hence, a symmetric many-particle state $|\psi\rangle=\sum_{i=1}^{N} \kappa_i|N/2,\xi_{\tiny{\rm ACS}}^{(i)}\rangle$ converges to $|\psi_N\rangle=\sum_{i=1}^{N} \kappa_i |\alpha^{(i)}\rangle$, where $|\alpha^{(i)}\rangle$ is a coherent state ($\alpha^{(i)}=\sqrt{N}\xi_{\rm ACS}^{(i)}$). Additionally, one can see that a many-particle state is entangled if there are more than one terms in the former expression. Similarly, a single-mode state is nonclassical if it is expressed as a superposition of more than one coherent states~\cite{vogel2014PRA}. Then, inseparability of $|\psi_N\rangle$ implies the nonclassicality of $|\psi\rangle$.

Therefore, one can adopt a many-particle inseparability criterion (for $N\to\infty$) to obtain a single-mode nonclassicality criterion. Ref.~\cite{tasgin2015HP} shows that spin squeezing criterion of Sorensen {\it et al.}~\cite{sorensen2001Nature} leads to quadrature squeezing condition~\cite{lee1991measure} for a single-mode field. This condition can be obtained by making Holstein-Primakoff (HP) transformation in the collective spin operators~\cite{emary2003chaos}, e.g. $\hat{S}_+\to\sqrt{N}\hat{b}$. A similar relation stands also for mixed states.

Due to the presence this intimate link between the three kinds of nonclassicalities, one can group the criteria into two~\cite{PScriteria}. (a) In the first group we can place: the spin squeezing criterion~\cite{sorensen2001Nature} for many-particle entanglement, quadrature squeezing condition for single-mode states~\cite{lee1991measure,ScullyZubairyBook}, and Duan-Giedke-Cirac-Zoller (DGCZ)~\cite{DGCZ_PRL2000} criterion (and its product form~\cite{Mancini&TombesiPRL2002_DGCZ_product}) with Simon-Peres-Horodecki (SPH)~\cite{SimonPRL2000,plenio2005logarithmic} criterion for two-mode entanglement. (b) The second group contains the Hillery-Zubairy (HZ) criterion~\cite{Hillery&ZubairyPRL2006} (which is a subset of conditions by Shchukin and Vogel~\cite{ShchukinVogelPRA2005}) for two-mode states, Mandel's Q parameter as the single-mode nonclassicality, and a many-particle criterion we still do not know yet. We note that sub-Poissonian criterion of Mandel's Q parameter can be obtained from the HZ criterion via BS method~\cite{hillery2006PRA,tasgin2015HP,tasgin2015measure}.

Group (a) is usually used for states generated from coherent (ACS or single-mode coherent) states via nonlinear hamiltonians~\cite{Kitagawa&UedaPRA1993,Tasgin&MeystrePRA2011} or squeezing transfer~\cite{PolzikPRL1999spinsqz}. DGCZ and SPH~\cite{SimonPRL2000} are necessary and sufficient criteria for Gaussian states~\cite{vitaliPRL2007optomechanical}. The second group (b) works better in witnessing the entanglement/nonclassicality of Fock-like single and two-mode states~\cite{NhaPRA2006Fock_states}. Common to both groups, it is possible to obtain stronger forms of two-mode criteria by using the Schr\"{o}dinger-Robertson inequality~\cite{Nha&ZubairyPRL2008} inplace of Heisenberg uncertainty version~\cite{Agarwal&BiswasNJP2005}.

In this paper, we aim to find the many-particle inseparability criterion missing in group (b). We try to {\it guess} its form from the single-mode  nonclassicality criterion which belongs to this group: namely Mandel's Q parameter (or sub-Poissonian distribution), i.e. $\langle (\hat{b}^\dagger \hat{b})^2\rangle-\langle \hat{b}^\dagger \hat{b} \rangle^2 < \langle \hat{b}^\dagger \hat{b}\rangle$. We try the simplest (not unique) way, $\sqrt{N}\hat{b}^\dagger \to \hat{S}_+$~\cite{emary2003chaos}. The question we start up with is simple. If we examine the uncertainty of $\hat{\mathcal{R}}=\hat{S}_+\hat{S}_-$, i.e. $\langle(\Delta \hat{\mathcal{R}})^2\rangle=\langle\hat{\mathcal{R}}^2\rangle-\langle\hat{\mathcal{R}}\rangle^2$, will we be able to obtain an inseparability criterion for many-particle systems?

This way, we obtain a criterion ($\xi_{\rm new}$) which works better than our expectations. The strength of violation of this criterion ($\xi_{\rm new}<0$, or larger squeezing in $\langle(\Delta\mathcal{R})^2\rangle$) accompanies the superradiant phase transition both for finite and infinite number of particles, see Fig.~\ref{fig2}. $\xi_{\rm new}$ also correctly predicts the temporal behavior of the entanglement of (timed) single-photon superradiance~\cite{scullyPRL2006timedDicke,svidzinskyPRA2008cooperative,svidzinskyPRA2010cooperative}, see Fig.~\ref{fig6}, for N=2000 atoms placed randomly in a sphere larger than a wavelength~\cite{svidzinskyPRA2008cooperative,svidzinskyPRA2010cooperative}. It is worth emphasizing that our derivation (also the validity) for $\xi_{\rm new}$ is completely independent of the presence of a relation between single-mode nonclassicality and many-particle inseparability.

We also obtain a criterion for ensemble-field entanglement, $\mu_{\rm new}<0$. We consider the stronger form [Eq.~(11) in Ref.~\cite{Nha&ZubairyPRL2008}] of the HZ criterion~\cite{Hillery&ZubairyPRL2006} for two-mode entanglement. We replace the operator $\hat{a}_1^\dagger\to\hat{S}_+$ for one of the two-modes, $\hat{a}_{1,2}$. 

(This has been performed in Ref.~\cite{RaymerPRA2003,PolzikNature2001ensemble,IgnacioPRA2011dissipatively} for the DGCZ criterion.) We observe that also $\mu_{\rm new}$ works very well for superradiant states, see Fig.s~\ref{fig3} and \ref{fig6}. Replacements $\hat{a}_1^\dagger\to\hat{S}_+$ and $\hat{a}_2^\dagger\to\hat{J}_+$ results in a criterion for ensemble-ensemble entanglement entanglement This works fine for detecting the entanglement between two ensembles after a Dicke-like phase transition~\cite{ZhengPRA2011ensemble_ensemble}. 

Wave function $\hat{\psi}(\bf r)$ becomes bunched (super-Poisonnian) above the critical atom-field coupling ($g>g_c$), see Fig.~\ref{fig4}(a). This is in contrast with the sub-Poissonian behavior of the collective (quasi-particle) excitations of the N-particle system (Fig.~\ref{fig2}(b)) and the scattered field (see Fig.~\ref{fig4}(b)). We remind that in the $N\to\infty$ limit collective excitations can be described by $\hat{b}$ operator alone. Such a behavior also occurs in the ground state of an interacting BEC (without a field). Bunching and many-particle entanglement emerge mutually when interaction (collisions) per particle exceeds the excitation energy, $U_{\rm int}/N > \hbar \omega_{\rm exc}$, see Fig.~\ref{fig5}. Incidentally, in experiments with BECs~\cite{Wright&BigelowPRA2008,Stamper&KetterlePRL1999phonons,Graham&WallsPRL1996scatering,tasginPRA2011vortex,Das2016NJPcollectively} we observe that BEC cannot recoil partially unless the excitation energy exceeds $U_{\rm int}/N$.

\subsection{Many-particle entanglement}

The derivation of $\mu_{\rm new}$ follows arguments similar to spin-squeezing condition by Sorensen {\it et al.}\cite{sorensen2001Nature}. Nevertheless, longer expressions show up due to the calculation of higher order moments. A many-particle system is separable if N-particle density matrix (DM) can be written in the form
\begin{equation}
\hat{\rho}=\sum_k P_k \: \rho_1^{(k)}\otimes\rho_2^{(k)}\otimes\ldots \otimes\rho_N^{(k)}\; ,
\label{DM}
\end{equation}  
where $\hat{\rho}_i^{(k)}$ is the DM of the $i^{\rm th}$ particle and $P_k$ is the classical probability for mixed states. Uncertainty of the $\hat{\mathcal{R}}=\hat{S}_+\hat{S}_-$ operator becomes larger than $\langle(\Delta\hat{\mathcal{R}})^2\rangle \geq \sum_k P_k (\langle\hat{\mathcal{R}}^2\rangle_k-\langle\hat{\mathcal{R}}\rangle_k^2)$ if we use the Cauchy-Schwartz inequality $\sum_k P_k \langle\hat{\mathcal{R}}\rangle^2 \geq (\sum_k P_k \langle\hat{\mathcal{R}}_k\rangle)^2$. We express the collective operators in terms of the single atom spins $\hat{s}_{\pm}^{(i)}$, e.g. $\hat{\mathcal{R}}=\hat{S}_+\hat{S}_-=\sum_{i_1=1}^N \sum_{i_2=1}^N \hat{s}_+^{(i_1)}\hat{s}_-^{(i_2)}$. We evaluate the difference $\langle\hat{\mathcal{R}}^2\rangle_k-\langle\hat{\mathcal{R}}\rangle_k^2$ using many Cauchy-Schwartz inequalities and relations among single particle operators, see the Supplementary Material~\cite{supplementary}. We show that the DM~(\ref{DM}) satisfies the inequality $\sum_k P_k \langle\hat{\mathcal{R}}^2\rangle_k-\sum_k P_k \langle\hat{\mathcal{R}}\rangle_k^2 \geq \eta_N$. We conclude that $\langle(\Delta\hat{\mathcal{R}})^2\rangle \geq \eta_N$ for a separable state. So, we define the parameter
\begin{equation}
\xi_{\rm new}=\langle(\Delta\hat{\mathcal{R}})^2\rangle_{\rho} - \eta_N
\end{equation}
whose negativity ($\xi_{\rm new}<0$) witnesses the inseparability of the many-particle system.

In Fig.~\ref{fig1}, we test $\xi_{\rm new}$ on Dicke states for N=16 particles (or S=8). $|S,m=\pm S \rangle$ states are separable, explicitly, $|g_1,g_2,\ldots g_N\rangle$ or $|e_1,e_2,\ldots e_N\rangle$ where $g_i$/$e_i$ means that the $i^{\rm th}$ particle is in the ground/excited state~\cite{MandelWolfbook,arecchi1972ACS}. The number of terms, hence the inseparability~\cite{}, increases upto $|S,m=0\rangle$. Linear entropy~\cite{brennen2003linearEntropy,meyer2001global,lambert2005entanglement}, an entanglement monotone~\cite{emary2004Qmonotone}, follows the expected result, such that it increases upto $|S,m=0\rangle$ state. Our criterion ($\xi_{\rm new}$) ---$\langle(\Delta\hat{\mathcal{R}})^2\rangle$ is more squeezed for more negative values of $\xi_{\rm new}$--- also follows the similar trend. Duan recently introduced a new criterion~\cite{duan2011many_particle_entanglement}, which not only serves for detecting the inseparability but it also reports that (if $\xi_{\rm Duan}>n$) at least $n$ number of particles are entangled. In Fig.~\ref{fig1}, we scaled $\xi_{\rm Duan}$ with the number 17. Hence, for $m=0$ it witnesses that at least 16 (all of the) particles are entangled. Duan's criterion is priceless in the research connecting the gravitation and entanglement~\cite{sonner2013holographic,jensen2013holographic,Susskind2013cool}, since it quantifies the depth (so the speed) of entanglement.   

\begin{figure}
\centering
\includegraphics[width=0.44\textwidth]{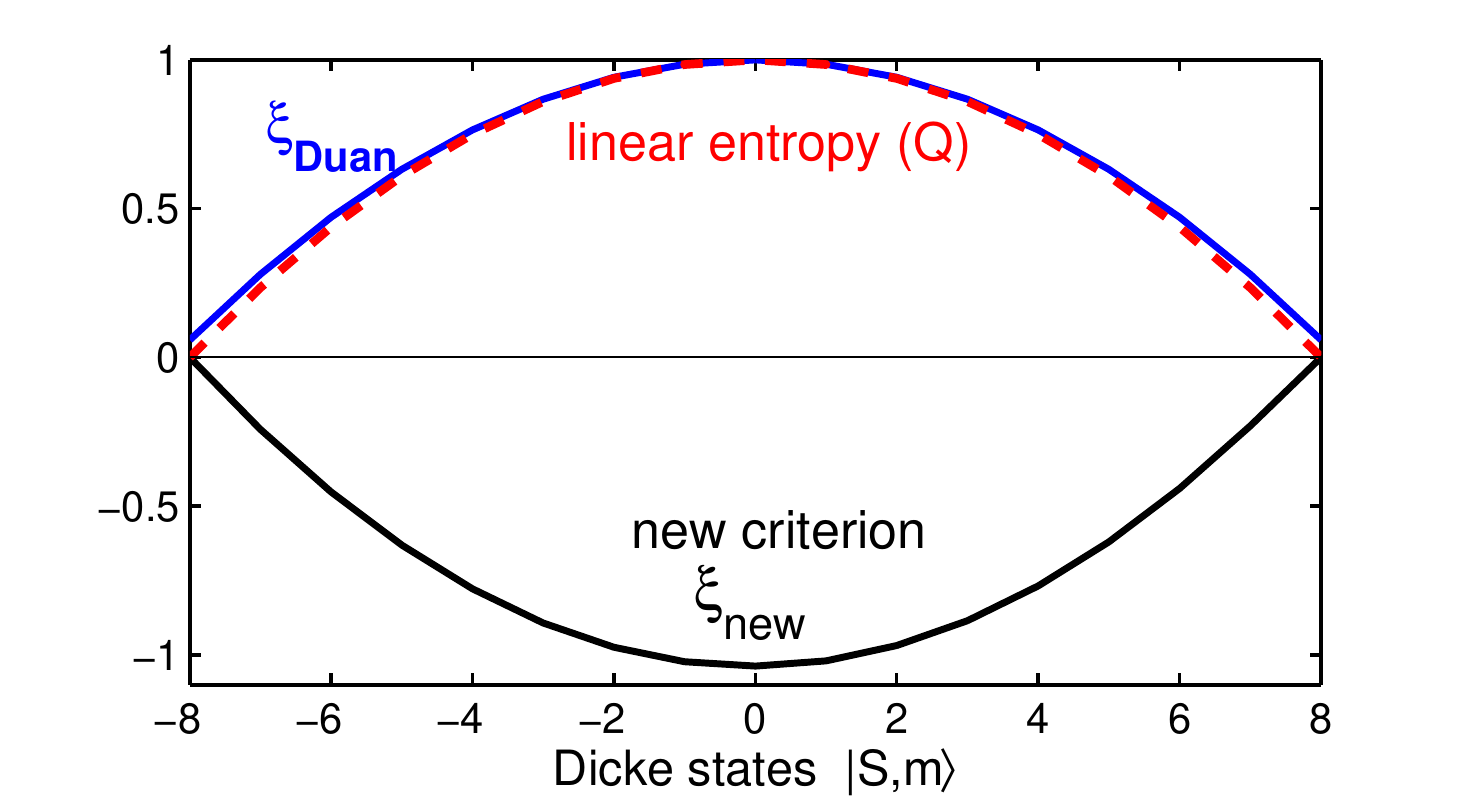}
\caption{\label{fig1} (color) The linear entropy Q~\cite{meyer2001global,lambert2005entanglement}, many-particle criterion by Duan~\cite{duan2011many_particle_entanglement} and the new many-particle criterion $\xi_{\rm new}$ successfully predicts the entanglement in Dicke states $|S,m\rangle$. $Q,\xi_{\rm Duan}>0$ and $\xi_{\rm new}<0$ implies entanglement.}
\label{fig1}
\end{figure}

In Fig.~\ref{fig2}, we calculate $\xi_{\rm new}$ for the ground state of the Dicke Hamiltonian
\begin{equation}
\hat{\mathcal{H}}=\hbar\omega_{eg} \hat{S}_z + \hbar\omega_a \hat{a}^\dagger\hat{a} + g/\sqrt{N}(\hat{S}_++\hat{S}_-)(\hat{a}^\dagger+\hat{a})
\label{DickeHamilt}
\end{equation}
in the thermodynamic limit ($N\to\infty$)~\cite{emary2003chaos} and simulate for symmetric subspace~\cite{Tasgin&MeystrePRA2011}. Here, $g$ is the atom-photon coupling strength where for $g>g_c=\sqrt{\omega_{eg}\omega_a}/2$ superradiant phase is observed~\cite{PS_RWA}.

$\xi_{\rm new}$ not only successfully predicts the presence of the many-particle inseparability, but also its negativity (squeezing in $\langle(\Delta\hat{\mathcal{R}})^2\rangle$) accompanies the order parameters ($\langle\hat{a}^\dagger\hat{a}\rangle$ and $\langle\hat{S}_z\rangle$) of the transition. In Fig.~\ref{fig2}(b), we observe that value of the linear entropy Q (an entanglement monotone~\cite{emary2004Qmonotone}) also accompanies the transition. The spin-squeezing criterion of Sorensen {\it et al.}~\cite{sorensen2001Nature} cannot witness the inseparability where $\xi_{\rm spin}<0$ implies the entanglement. The criterion of Duan~\cite{duan2011many_particle_entanglement} (not plotted in Fig.~\ref{fig2}(b)) does not exceed 1 for the ground state, which is a superposition of many Dicke states.

\begin{figure}
\centering
\includegraphics[width=0.44\textwidth]{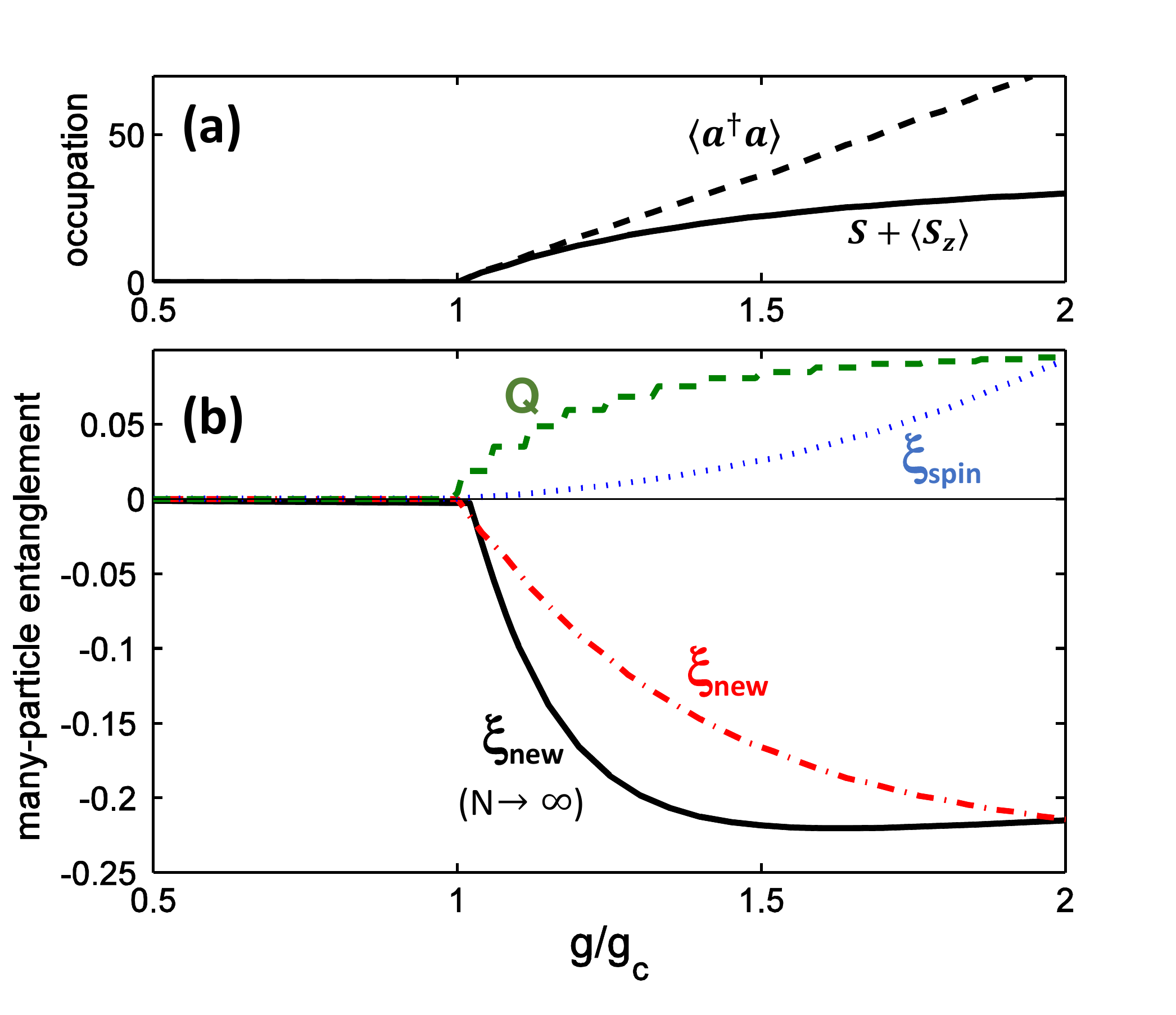}
\caption{\label{fig2} (color) (a) The number of photons and the excitation of the ensemble in the ground state of Dicke hamiltonian~\ref{DickeHamilt}. Above the critical atom-photon coupling strength, $g>g_c$, superradiant phase transition occurs. (b) Many-particle entanglement. Linear entropy Q and the new criterion $\xi_{\rm new}$ (squeezing in $\langle(\Delta\hat{\mathcal{R}})^2\rangle$) accompanies the order parameters of the phase transition. $Q>0$ and $\xi_{\rm spin},\xi_{\rm new}<0$ implies entanglement.}
\label{fig2}
\end{figure}

In both Fig.~\ref{fig1} and Fig.~\ref{fig2}(b), linear entropy Q and $\xi_{\rm new}$ exhibit parallel behavior. So, we became curious if this is true also for random states. For N=16, in the $2^{16}$ dimensional space, we generated random states and examined if $\xi_{\rm new}$ and Q display parallel behavior. Even though $\xi_{\rm new}$ managed to detect the inseparability of all 2000 states, when $Q>0$, the two did not exhibit parallel behavior always.

\subsection{A stronger single-mode nonclassicality criterion}

 Atomic coherent states (ACSs) are the many-particle states in the symmetric subset if the complete set of Dicke states~\cite{dicke1954SR}. In the limit $N\to\infty$, ACSs converges to coherent states of light~\cite{radcliffe1971JPhysA,klauder1985applications}. A many-particle (single-mode) state is inseparable (classical) if it is the superposition of more than one ACSs (coherent state). Therefore, a many-particle criterion converges to a criterion for single-mode nonclassicality. Alternatively, one can perform Holstein-Promakoff transformation, e.g. $\hat{S}_+=\hat{b}^\dagger\sqrt{N-\hat{b}^\dagger}\hat{b}$, and let the limit $N\to\infty$.

We now have a many-particle criterion in group (b) involving $\hat{S}_{\pm,z}$ collective operators. Therefore, we check if we obtain the same (or similar) single-mode criterion, that is Mandel's Q parameter, in the $N\to\infty$ limit. We obtain the criterion (see the Supplementary Material)
\begin{equation}
\langle(\Delta \hat{n})^2\rangle \geq \langle\hat{n}\rangle + \big(\operatorname{Im}\{\langle\hat{b}^{\dagger}{}^2\rangle-\langle\hat{b}\rangle^2 \}\big) ^2 \: ,
\label{singlemode_new}
\end{equation}  
where $\hat{n}=\hat{b}^\dagger\hat{b}$. We note that this is sub-Poissonian criterion except the last term. When one rotates the coordinates~\cite{ScullyZubairyBook} $\hat{b}_{\theta}=\hat{b}e^{i\theta}$, the last term becomes $\operatorname{Im}\{(\langle\hat{b}^{\dagger}{}^2\rangle-\langle\hat{b}\rangle^2)e^{i2\theta} \}$ which is equal to zero for the proper choice of the phase $e^{i2\theta}$. In this situation, we recover the Mandel's Q parameter. 

In general, however, criterion~(\ref{singlemode_new}) seems to be a hybrid [both group (a) and (b)] one. From our previous experience~\cite{ge2015conservation} we know that the last term becomes maximum, for Gaussian states, when the $\theta$ is chosen in the direction of maximum quadrature squeezing. Since the first two terms are independent from rotations one can make the test stronger via rotations.

A question we need to answer is the following. We started with a weaker criterion (sub-Poissonian), but obtained a stronger one, (\ref{singlemode_new}), on return? The answer is the following. We only calculated the uncertainty $\Delta(\hat{S}_+\hat{S}_-)$ being inspired from the uncertainty $\Delta(\hat{b}^\dagger\hat{b})$. We did not transform the right hand side ($\langle\hat{b}^\dagger\hat{b}\rangle$) of the weaker inequality.

\subsection{Ensemble-field entanglement}

Commonly used two-mode criteria can be put in a stronger form using the Schr\"odinger-Robertson inequality and the partial tranpose of the operator~\cite{Nha&ZubairyPRL2008}. For instance, the product form of the DGCZ criterion~\cite{Mancini&TombesiPRL2002_DGCZ_product,Agarwal&BiswasNJP2005}, belonging to group (a), can be put in a stronger form by using the variances $\tilde{H}_1'=\hat{x}_1+\hat{x}_2$ and $\tilde{H}_2'=\hat{p}_1=\hat{p}_2$ in the Schr\"odinger-Robertson inequality~\cite{Nha&ZubairyPRL2008}. Similarly, a stronger form of the HZ criterion, in group (b), can be obtained using the $\tilde{H}_1=\hat{a}_1^\dagger\hat{a}_2+\hat{a}_1^\dagger\hat{a}_2$ and $\tilde{H}_2=i(\hat{a}_1^\dagger\hat{a}_2-\hat{a}_2^\dagger\hat{a}_1)$ in the Schr\"odiger-Robertson inequality, see Eq.~(11) in Ref.~\cite{Nha&ZubairyPRL2008}.

Ref.s~\cite{RaymerPRA2003,PolzikNature2001ensemble,IgnacioPRA2011dissipatively} show that it is possible to obtain a criterion for the ensemble-field entanglement by making the substitutions $\hat{x}_1\to \hat{S}_x$ and $\hat{p}_1\to \hat{S}_y$ in $\tilde{H}_{1,2}'$. Similar to Ref.s~\cite{RaymerPRA2003,PolzikNature2001ensemble,IgnacioPRA2011dissipatively}, we perform the substitutions $\hat{a}_1^\dagger\to \hat{S}_+$ and $\hat{a}_1\to\hat{S}_-$ in $\tilde{H}_{1,2}$,
\begin{equation}
\tilde{H}_1=\hat{S}_+\hat{a}_2+\hat{S}_-\hat{a}_2^\dagger \quad {\rm and} \quad  
\tilde{H}_2=i(\hat{S}_+\hat{a}_2-\hat{S}_-\hat{a}_2^\dagger) \: ,
\end{equation}
and obtain the parameter
\begin{eqnarray}
\mu_{\rm new}^{\rm SR}=\big(  \langle (\Delta \tilde{H}_1)^2\rangle -2\langle\hat{S}_z\rangle \big) \:
\big(  \langle (\Delta \tilde{H}_2)^2\rangle -2\langle\hat{S}_z\rangle \big) \nonumber
\\
-| \langle -\hat{S}_+\hat{S}_- + 2\hat{S}_z\hat{a}\hat{a}^\dagger \rangle |^2 
-\langle \Delta\tilde{H}_1\Delta\hat{H}_2\rangle_{\rm s}^2 \: ,
\label{muHZSR}
\end{eqnarray}
where $\mu_{\rm new}^{\rm SR}<0$ witnesses the presence of the ensemble-field entanglement.

In Fig.~\ref{fig3}, we plot $\mu_{\rm new}^{\rm SR}$ for finite/infinite number of particles. We observe that violation of $\mu_{\rm new}^{\rm SR}$, squeezing in the product (\ref{muHZSR}), accompanies the order parameters given in Fig.~\ref{fig2}(a). For the purposes of comparison, we also calculate  $\mu_{\rm new}^{\rm HZ}$. We performe the substitution $\hat{a}_1^\dagger \to \hat{S}_+$ in the HZ criterion~\cite{Hillery&ZubairyPRL2006}, $\langle\hat{a}_1^2\hat{a}_2^\dagger{}^2\rangle < \langle\hat{a}_1^\dagger{}^2\hat{a}_1^2 \hat{a}_2^\dagger{}^2\hat{a}_2^2\rangle$, which is weaker than Ref.~\cite{Nha&ZubairyPRL2008}. In Fig.~\ref{fig3}, we see that $\mu_{\rm new}^{\rm HZ}$ cannot witness the ensemble-field entanglement for $g>1.9g_c$. The spin-squeezing criterion~\cite{sorensen2001Nature} $\mu_{\rm spin}$ cannot reveal the presence of entanglement at all.

\begin{figure}
\centering
\includegraphics[width=0.44\textwidth]{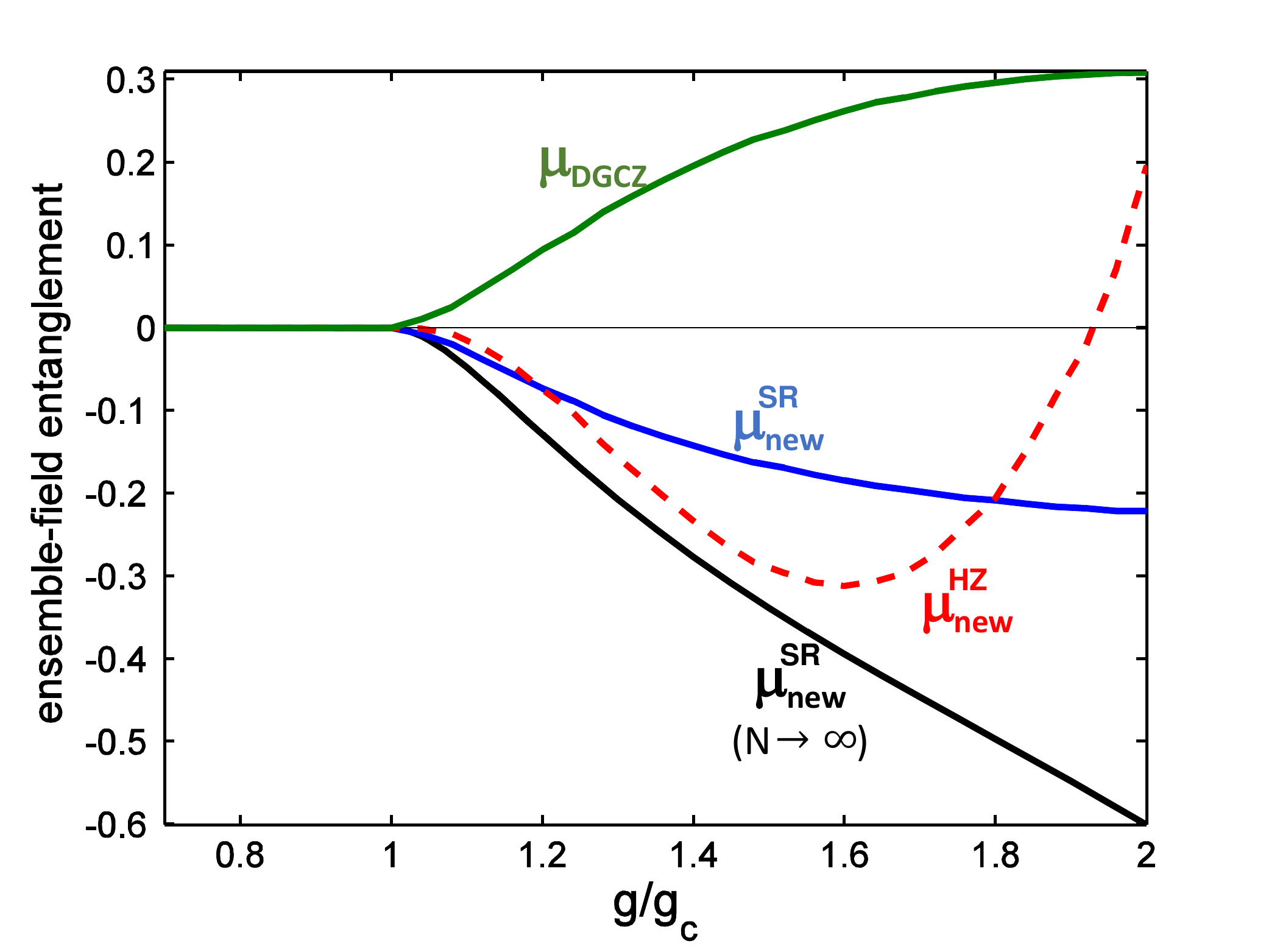}
\caption{\label{fig3} (color) Ensemble-field entanglement in the ground state of the Dicke hamiltonian~(\ref{DickeHamilt}) for finite/infinite number of particles. $\mu<0$ witnesses the entanglement. $\mu_{\rm new}^{\rm HZ}$ is obtained from the HZ~\cite{Hillery&ZubairyPRL2006} criterion. $\mu_{\rm new}^{\rm SR}$ is obtained from the stronger form of the HZ criterion~\cite{Nha&ZubairyPRL2008}.}
\label{fig3}
\end{figure}

\subsection{Bunching in the wave function}
 Superradiant scattering from a Bose-Einstein condensate (BEC) has been studied extensively in the last two decades.  In the cases of directional scattering~\cite{Ketterle1999Science,MeystrePRL1999_BEC_SR} or scattering into a cavity~\cite{EsslingerNature2010Dicke,Nagy&DomokosPRL2010Dicke}, wave function operator can be expressed into two modes $\hat{\psi}({ \bf r})=u_g({ \bf r})\hat{c}_g + u_e({ \bf r}) \hat{c}_e$. Here, $u_e({ \bf r})=e^{i{\bf k}\cdot {\bf r}}u_g({ \bf r})$ and $u_e({\bf r})=\cos(kx)u_g({ \bf r})$ for the two cases, respectively. Hence, one can calculate the bunching of the atoms in the condensate, $g^{(2)}=\langle \hat{\psi}^\dagger({\bf r})\hat{\psi}^\dagger({\bf r})\hat{\psi}({\bf r})\hat{\psi}({\bf r})\rangle$, to learn how the other atoms react to the measurement (modification) of a single one. In Fig.~\ref{fig4}, we show that atoms display bunched behavior above the phase transition.

This behavior is opposite to the one for collective excitations ($\hat{S}_+\to\hat{c}_e^\dagger\hat{c}_e\to\sqrt{N}\hat{b}$ in the limit $N\to\infty$) and the scattered field $\hat{a}$. $\xi_{\rm new}<0$ implies that $\hat{b}$ (quasi-particle) field is anti-bunched in this limit. In Fig.~\ref{fig4}(b), we observe the anti-bunching in the scattered field $\hat{a}$.

\begin{figure}
\centering
\includegraphics[width=0.44\textwidth]{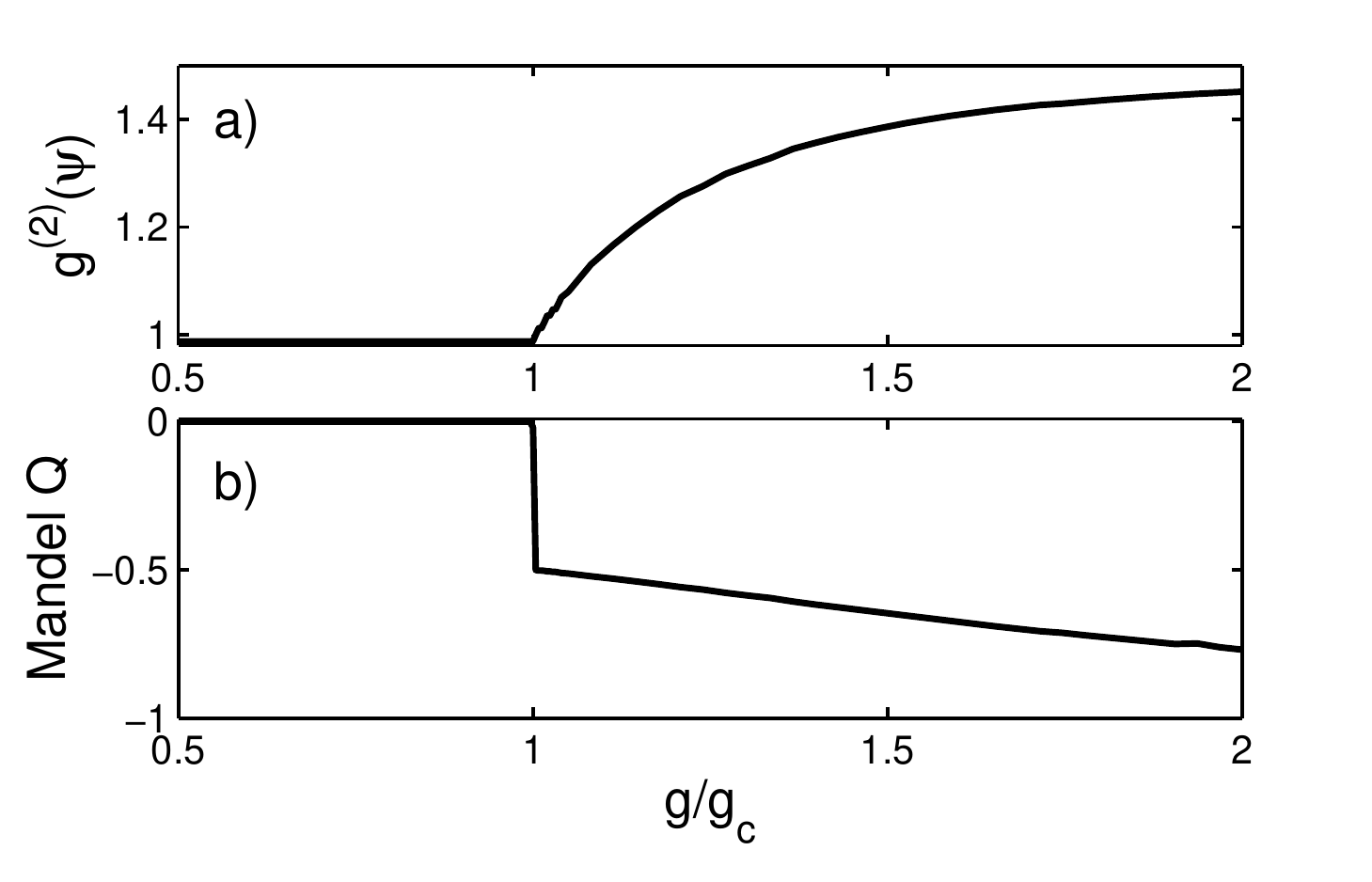}
\caption{\label{fig4} (a) Wave-function operator, describing indistinguishable atoms in a BEC, is bunched in the superradiant phase. $g^{(3)}$ and  $g^{(4)}$ display behavior similar to $g^{(2)}$. (b) In contrast to the wave function, the scattered field $\hat{a}$ displays sub-Poissonian (anti-bunching) behavior.}
\label{fig4}
\end{figure}

Experiments on BECs~\cite{Wright&BigelowPRA2008,Stamper&KetterlePRL1999phonons,Graham&WallsPRL1996scatering,tasginPRA2011vortex,Das2016NJPcollectively} show that a condensate reacts to an excitation collectively unless the energy of the excitation $\hbar\omega_{\rm exc}$ exceeds the interaction energy per atoms $U_{\rm int}/N=g_s\int d^3{\bf r} |\psi({\bf r})|^4/N$, where $g_s$ is the strength of collisions. If $\hbar\omega_{\rm exc}>U_{\rm int}$, we observe that atoms in the condensate can be recoiled partially. This phenomenon made us raise the question "what happens in the ground state of a stand alone (no field) interacting BEC ?". The hamiltonian
{\thinmuskip=2mu
\medmuskip=3mu plus 2mu minus 3mu
\thickmuskip=4mu plus 5mu minus 2mu
\begin{equation}
\hat{\mathcal{H}}=\int d^3{\bf r}\hat{\psi}({\bf r})^\dagger\hat{H}_0({\bf r}) \hat{\psi}({\bf r}) 
+ g_s \int d^3{\bf r} \hat{\psi}({\bf r})^\dagger \hat{\psi}({\bf r})^\dagger \hat{\psi}({\bf r}) \hat{\psi}({\bf r})
\label{H_BEC1}
\end{equation}}
transforms to 
\begin{equation}
\hat{\mathcal{H}}=\hbar\omega_{\rm exc}\hat{S}_z + U_{\rm int} \hat{S}_z^2
\label{H_BEC2}
\end{equation}
similar to Ref.~\cite{sorensen2001Nature}. Here, we assume that $\omega_{\rm exc}$ is the excitation of the BEC to a higher energy level. For a BEC, in a harmonic trap, harmonic oscillator spacing ($\sim$100 Hz) is much smaller than the kinetic energy BEC gains due to recoil ($\sim {\rm 10}^4{\rm -{\rm 10}^5}$Hz)~\cite{Ketterle1999Science}. We examine the ground state of this system and find an interesting coincidence. The GS of the BEC becomes many-particle entangled and wave-function becomes bunched after $U_{\rm int}>\hbar\omega_{\rm exc}$.

\begin{figure}
\centering
\includegraphics[width=0.47\textwidth]{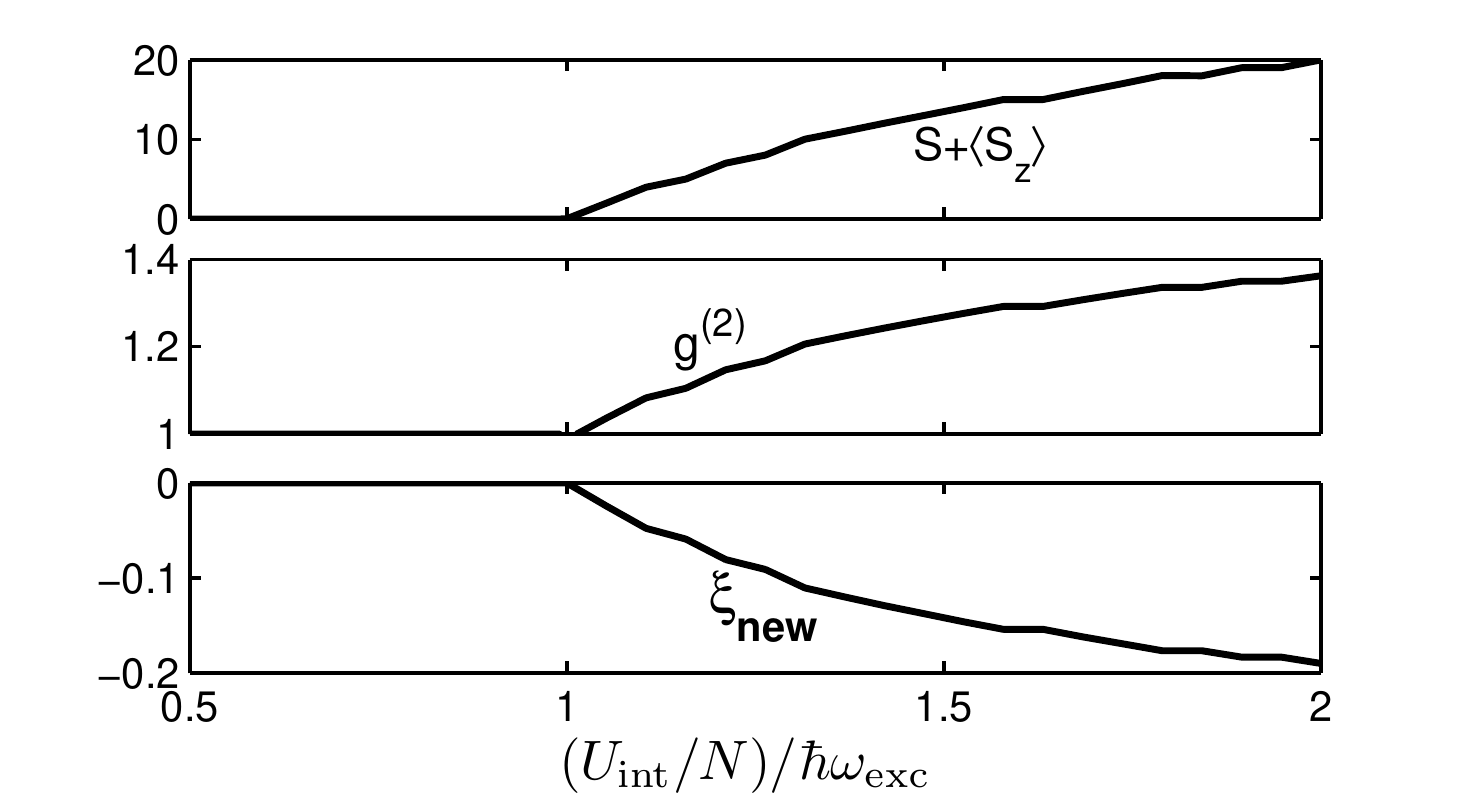}
\caption{\label{fig5} The ground state of an interacting BEC, hamiltonian (\ref{H_BEC1}) or (\ref{H_BEC2}), when there is no field present. When $U_{\rm int}/N > \hbar\omega_{\rm exc}$, the ground state of the BEC becomes many-particle entangled ($\xi_{\rm new}<0$) as well as atoms are bunched ($g^{(2)}>1$). Incidentally, experiments with BECs~\cite{Wright&BigelowPRA2008,Stamper&KetterlePRL1999phonons,Graham&WallsPRL1996scatering,tasginPRA2011vortex,Das2016NJPcollectively} show that BEC responses the external excitations collectively unless an $\hbar\omega_{\rm exc}> U_{\rm int}/N$ is transferred to a single atom.}
\label{fig5}
\end{figure}

\subsection{ Single-photon superradiance}

 Single-photon superradiance is one of the few (almost) exactly solvable many-body systems~\cite{svidzinskyPRA2008cooperative,svidzinskyPRA2010cooperative} and it is gaining importance due to its technological applications~\cite{ScullyPRL2015Subradiance,ScullyScience2009SR}. Temporal behavior of a timed Dicke state~\cite{scullyPRL2006timedDicke}, prepared initially in the state $|\psi(0)\rangle=\sum_{j=1}^N e^{i{\bf k}_0\cdot {\bf r}_j}|g_1,g_2,\ldots e_j,\ldots,g_N\rangle$, can be given as~\cite{svidzinskyPRA2008cooperative,svidzinskyPRA2010cooperative}
{\thinmuskip=2mu
\medmuskip=3mu plus 2mu minus 3mu
\thickmuskip=4mu plus 5mu minus 2mu
\begin{equation}
|\psi(t)\rangle = \sum_{j=1}^N \beta_j(t)|g_1 \dots e_j \ldots g_N\rangle |0\rangle + \sum_{\bf k} \gamma_{\bf k}(t) |g_1 \ldots g_N \rangle |1_{\bf k} \rangle
\label{psit}
\end{equation}}
The solutions of $\beta_j(t)$ and $\gamma_{\bf k}(t)$ are studied in Ref.s~\cite{svidzinskyPRA2008cooperative,svidzinskyPRA2010cooperative} intensively. We  test our criteria $\xi_{\rm new}$ and $\mu_{\rm new}$ also for the single-photon superradiance of 2000 atoms randomly placed at positions ${\bf r}_j$. The spatial extent of the ensemble is 5 times larger than the wavelength $\lambda_0=2\pi/k_0$. 

In Fig.~\ref{fig6}, we observe that the initial many-particle entanglement is lost after $t>1/\Gamma_N$, where collective decay rate $\Gamma_N\sim N\gamma$ 
can be much larger than the single atom decay rate $\gamma$. This is something expected from Eq.~(\ref{psit}), since the particles decay to the separable state, where $\beta_j(t)~e^{-\Gamma_N t}$~\cite{svidzinskyPRA2008cooperative,svidzinskyPRA2010cooperative}. We also examine the entanglement of the ensemble with the central mode (${\bf k}_0$). Initially $\mu_{\rm new}=0$ since $\gamma_{{\bf k}}(0)=0$. For $t>0$, $\mu_{\rm new}$ witnesses the inseparability as $\beta_j(t)$ and $\gamma_{\bf k}(t)$ are mixed in $|\psi(t)\rangle$. Finally, $\mu_{\rm new}$ approaches to zero again since the system ends up with the $\gamma_{\bf k}$ states eventually.

\begin{figure}
\centering
\includegraphics[width=0.44\textwidth]{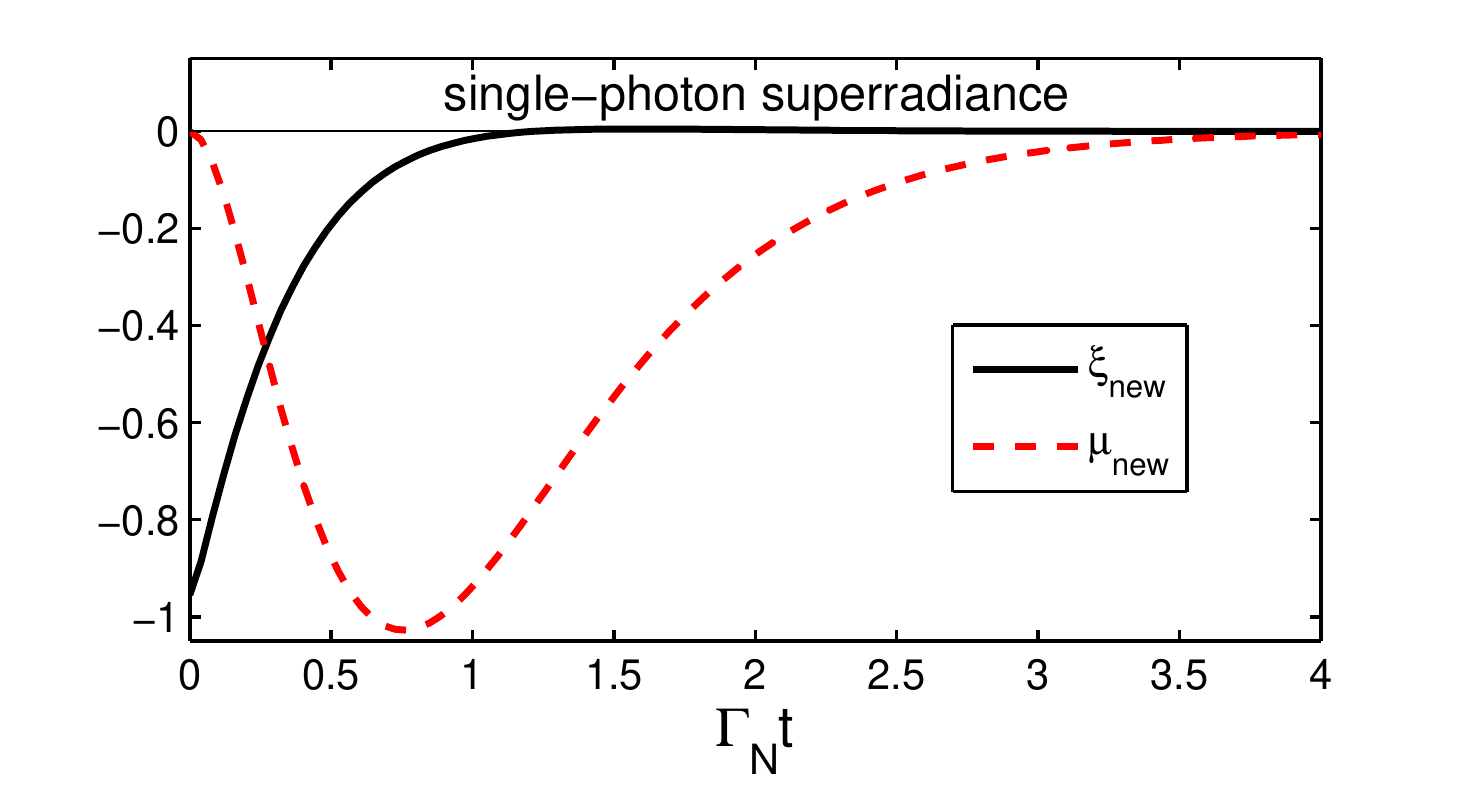}
\caption{\label{fig6} (color) Temporal behavior of  many-particle entanglement ($\xi_{\rm new}$) and ensemble-field entanglement ($\mu_{\rm new}$) for  single-photon superradiance of N=2000 atoms placed randomly in a sphere larger than wavelength. }
\label{fig6}
\end{figure}

Finally, we anticipate that derivations, $\langle(\Delta\hat{\mathcal{R}})^2 \rangle$, leading to $\xi_{\rm new}$ can be utilized for calculating the entanglement depth~\cite{duan2011many_particle_entanglement} of the system. This can be carried out by grouping the inseparable particles in the density matrix (see Eq.s~(1) and (2) in Ref.~\cite{duan2011many_particle_entanglement}), in place of using a full-separable density matrix given in Eq.~(\ref{DM}).

\subsection{Acknowledgments}
\begin{acknowledgements}
I gratefully thank M. Suhail Zubairy for his hospitality in Texas A\&M University and for participating in the 1st Quantum Optics Workshop in Turkey. He did not cancel his talk even after the attack (in the same city) where more than hundred people lost their lives. Part of this research has been initiated after Moochan (Barnabas) Kim raised the question ``Can we distinguish between different single-photon Dicke states according to their entanglement strength?". I thank Anatoly A. Svidzinsky for his help on single-photon superradiance. I thank Marlan O. Scully for encouraging me to raise questions in the meetings and the talks. 

\end{acknowledgements}

\bibliography{bibliography}

\begin{thebibliography}{63}
\bibitem [{PS_()}]{PS_RWA}%
  \BibitemOpen
  \href@noop {} {}\bibinfo {note} {We obtain the similar results when rotating
  wave approximation applied on the
  hamiltonian~(\ref{DickeHamilt}).}\BibitemShut {Stop}%
\end{thebibliography}

%
%
%
%
%

\end{document}